# Vortex Quasi-Crystals

C. A. R. Sá de Melo

*School of Physics, Georgia Institute of Technology, Atlanta, Georgia 30332*
(August 01, 1999)

There seems to be a one to one correspondence between the phases of atomic and molecular matter (AMOM) and vortex matter (VM) in superconductors. Crystals, liquids and glasses have been experimentally observed in both AMOM and VM. However, quasi-crystals also exist in AMOM, thus a new phase of vortex matter is proposed here: the vortex quasi-crystal. It is argued that vortex quasi-crystals are stabilized due to boundary and surface energy effects for samples of special shapes and sizes, and that a phase transition between a vortex crystal and a vortex quasi-crystal occurs as a function of magnetic field and temperature as the sample size is reduced.

The general subject of vortex physics in superconductors is quite interesting since there seems to be a large variety of possible equilibrium vortex phases in superconductors (layered and not layered) as pointed out in the very interesting article by Crabtree and Nelson [1]. The term "vortex matter" has been coined to emphasize the complexity and diversity of vortex phases in superconductors when compared to atomic and molecular matter. One can think of a one to one correspondense between phases in atomic and molecular matter (AMOM) and phases in vortex matter (VM). A liquid in AMOM corresponds to a vortex liquid in VM; a crystalline lattice in AMOM corresponds to a vortex lattice in VM; an amorphous or glassy solid in AMOM corresponds to an amorphous or glassy vortex system in VM. However, among all the possibilities discussed in the vortex matter literature, a very interesting one is missing: a vortex quasi-crystal. Quasi-crystals in AMOM where experimentally discovered several years ago [2], and they were classified in the nice review article by Mermin [3]. Thus, the present paper is dedicated to the first studies of vortex quasi-crystalline phases.

In their interesting paper Crabtree and Nelson [1] discuss that the detailed equilibrium behavior of vortex phases depends essentially on the competition of four energies: a) the thermal energy favors a vortex liquid of lines or pancakes; b) the vortex interaction energy favors a perfect lattice; c) the pinning energy favors an amorphous or glassy solid; d) the coupling energy between layers (in layered superconductors) controls the formation of vortex lines from weakly interacting pancake vortices in adjacent layers. When these four energies are about the same order of magnitude, an unprecedented variety of liquid and solid phases and transitions among them appear as magnetic field, temperature, and applied current are varied. Their analysis is quite generally true for bulk materials and bulk vortex matter properties, where surface energies are not important. It will be argued here however, that vortex quasi-crystals maybe stabilized by boundary effects and surface energies.

The central question of this manuscript is: are vortex quasi-crystals new phases of vortex matter? If so, what suggestions can be given to experimentalists to help in the search for such phases? In order not to delude the reader, it must be emphasized that these questions will be addressed here on a preliminary basis, and further detalied work will be necessary. In order to consider the possibility of a vortex quasi-crystal one must seek under what conditions a quasi-crystalline arrangement is at all possible. A definite possibility is to argue that a stable vortex quasi-crystal arises from boundary effects in samples, where the sample size and shape play an important role in the vortex quasi-crystal stabilization. For instance, pentagon cylinder or decagon cylinder samples may be good candidates to produce 5-fold or 10-fold vortex-quasicrystals. If boundaries are important then in the thermodynamic limit of infinite volume it is generally difficult to produce vortex quasi-crystalline order.

This possibility will be considered here under the following program. In this manuscript only the case of a three dimensional isotropic type II superconductor in a magnetic field is considered. First, the bulk free energy is calculated for a triangular, square and a 5-fold quasi-crystal array. The 5-fold quasi-crystal array is modeled by a Penrose tiling of the plane. It is shown that the Penrose tile array (vortex quasi-crystal) has a bulk free energy which is just a few percent higher than the triangular array. Second, instead of considering an infinite (bulk) system, a decagon cylinder sample is discussed. The decagon cylinder has a decagonal cross-section (in the xy plane) with side dimension $\ell$, and with height $L \gg \ell$ along the z direction. In this case, when the sample size gets smaller the contribution of the boundaries (surface energy) to the total free energy of the system becomes more important. The surface energy is highly sensitive to the symmetry and to the surface area of the boundaries. Taking into account the surface free energies, it is shown that the Penrose tile (vortex quasi-crystal) has lower free energy than the triangular lattice in certain regions of the magnetic field versus temperature phase diagram. This is suggestive that a "first order phase transition" may occur between the triangular lattice and the Penrose tiling (vortex quasi-crystal). [4]

For simplicity, the starting point of the analysis to follow is the Ginzburg-Landau free energy density

$$\Delta F_s = F_1 + \frac{1}{2m}\left|\left(-i\hbar\nabla - \frac{2e\mathbf{A}}{c}\right)\Psi(\mathbf{r})\right|^2 + \frac{H^2}{8\pi} \quad (1)$$

for a bulk isotropic superconductor with no disorder, where $\Delta F_s = F_s - F_n$ is the free energy diffence between the superconducting state ($F_s$) and the normal state ($F_n$), and $F_1 = \alpha|\Psi(\mathbf{r})|^2 + \beta|\Psi(\mathbf{r})|^4/2$. It is useful, however, to introduce the dimensionless quantities $\Psi = \Psi_0 f$, $\rho = \mathbf{r}/\lambda(T)$, $\mathcal{A} = \xi(T)\mathbf{A}/\phi_0$, and $\mathcal{H} = \xi(T)\lambda(T)|\mathbf{H}|/\phi_0$. Here, $f = f_0 \exp(i\phi)$, and $\mathbf{A} = \mathbf{A}_0 + \nabla\phi/\kappa$. Notice that when $f_0 = 1$ the system is fully superconducting, and when $f_0 = 0$ the system is normal, thus $f_0 \leq 1$ always. Considering the minimization of free energy difference with respect to $\mathcal{A}$ and $\Psi$ it is easy to arrive at equations for the dimensionless functions $\mathbf{H}$ and $f_0$. The microscopic field is

$$\mathcal{H}(x,y) = \kappa\left[1 + \frac{H - H_{c_2}}{H_{c_2}}\right] - \frac{g(x,y)}{2\kappa}, \quad (2)$$

where $\mathbf{H}$ is parallel to the $z$ direction, i.e., $\mathbf{H} = \mathcal{H}\hat{\mathbf{z}}$, and the equation for $f_0$ can be written as

$$\nabla^2(\log g) + 2\kappa^2 = 0, \quad (3)$$

where $g = f_0^2$ is a positive definite function of x and y, i.e., $g(x,y) \geq 0$. Notice that $\mathcal{H} = \kappa$ ($H = H_{c_2}$) for $f_0 = 0$ ($g = 0$). The most general solution of Eq. (3) has the form

$$g(x,y) = \exp\left[-\kappa^2(x^2 + y^2)/2\right] \Xi(x,y), \quad (4)$$

where $\Xi(x,y) = \exp[\gamma(x,y)]$ and $\gamma(x,y)$ satisfies Poisson's equation, i.e., $\nabla^2\gamma(x,y) = 0$. This means that $\gamma(x,y)$ is a harmonic function and can be expressed as the real part of **any** analytical function of $z = x + iy$. This observation has very important consequences for the microscopic field profile $\mathcal{H}(x,y)$ of eqn. (2), which depends strongly on the structure of $g(x,y)$.

The bulk Gibbs free energy can be easily expressed in dimensionless units as [5]

$$G_s(H,T) = G_n(H,T) - \frac{(\kappa - H)^2}{(2\kappa^2 - 1)\beta}, \quad (5)$$

where the parameter $\beta = \langle g^2 \rangle / \langle g \rangle^2$ is a geometrical factor independent of $\kappa$. The notation $\langle \cdots \rangle$ indicates average over volume. It is important to notice that $\beta \geq 1$ no matter what is the form of $g(x,y)$ because of the Schwartz inequality. In addition, notice that the Gibbs free energy above is a minimum, whenever $\beta$ reaches its minimum value. Furthermore, notice that the mixed state has lower Gibbs free energy than the normal state for $H < H_{c_2}$.

For the purpose of calculating the parameter $\beta$ and the free energies corresponding to different vortex configurations, the analytical structure of $g(x,y)$ in the complex plane is used to rewrite it as

$$g(z,\bar{z}) = exp(-\kappa z\bar{z}/2)|P(z)|, \quad (6)$$

where $P(z) = \mathcal{N}\prod_{i=1}^{M}(z - z_i)$. Here each $z_i$ corresponds to a zero of the order parameter in the complex plane, and $M$ is the number of zeros. The zeros $z_i$ indicate the location of vortices. From now on it is assumed that there is only one vortex with flux $\Phi_0$ at each position $z_i$, i.e., each zero in non-degenerate. In this case, M corresponds to the number of vortices, and thus the total flux threading the sample is $\Phi = M\Phi_0$. The normalization coefficient $\mathcal{N}$ just guarantes that $g(z,\bar{z}) \leq 1$. Depending on the locations of the zeros $g(z,\bar{z})$ it is possible to study several possibilities of periodic and quasi-periodic vortex arrangements. In this study though, only vortex crystals corresponding to triangular and square lattices and vortex quasi-crystals corresponding to the 5-fold Penrose tiling of the plane will be discussed. Both the square lattices and triangular lattices can be generated via the tiling method, i.e., via the periodic arrangements of identical square tiles or identical lozenges of internal angles ($60^o$ and $120^o$). The Penrose lattice, however, requires quasi-periodic arrangements of **two** types of tiles (lozenges), one with internal angles $36^o$ and $144^o$ and the other with internal angles $72^o$ and $144^o$. Using the representation in eqn. 6, the values of $\beta$ for the triangular, square and Penrose tiling are respectively $\beta_3 = 1.16$, $\beta_4 = 1.18$ and $\beta_5 = 1.22$. This immediately indicates that the triangular lattice has lower free than the square lattice which has lower free than the 5-fold vortex quasi-crystal (Penrose tiling). The order of the free energies should not come out as a surprise, however the fact that the free-energy difference between the triangular and 5-fold vortex quasi-crystal is so low suggests that appropriate boundaries can favor 5-fold symmetry as the sample size gets smaller.

In order to investigate how boundary effects can modify the total free energy of the system, a sample in the shape of a decagon cylinder of side $\ell$ is considered. Imposing that no currents flow through the sample boundaries leads to the condition $\hat{\mathbf{n}} \cdot [\nabla/i - 2\pi\mathbf{A}/\phi_0] \psi(x,y) = 0$ in all ten boundaries. The unit vector $\hat{\mathbf{n}}$ points along the normal direction of each facet of the decagon cylinder. The boundary conditions can be translated in terms of the harmonic function $\gamma(x,y)$ as

$$n_x \frac{\partial\gamma}{\partial y} - n_y \frac{\partial\gamma}{\partial x} = \kappa(n_x y - n_y x), \quad (7)$$

where $n_x$ and $n_y$ are just the x and y components of the normal unit vector $\hat{\mathbf{n}}$ at each one of the decagon cylinder side faces. The solution for this boundary value problem can be obtained using a Schwarz-Christoffel conformal map of the decagon to a semi-infinite plane

$$\frac{dz}{dw} = A \prod_j (w^2 - x_j^2)^{-1/5} \quad (8)$$

where the vertices of the decagon located at $z_i$ are mapped into the points $(\pm x_1, \pm x_2, \pm x_3, \pm x_4, \pm x_5)$ in the real axis of the $w$-plane. The full solution of this problem is quite complicated, and requires heavy use of numerical methods. [6] However, the additional surface free energy can be estimated due to the boundary mismatch of the bulk solution (interior solution) with respect to the solution near the surfaces (exterior solution). The Gibbs free energy difference $\Delta G = G_3 - G_5$ (in dimensional units) between the triangular and the 5-fold quasi-periodic Penrose structure then becomes

$$\Delta G = -\frac{1}{8\pi}\frac{(H_{c_2}-H)^2}{(2\kappa^2-1)(\beta^*)^2} + \frac{H_c^2}{4\pi}\frac{(\alpha_3-\alpha_5)}{R_e}, \qquad (9)$$

where $\beta^* = \sqrt{\beta_3\beta_5/(\beta_5-\beta_3)}$, $\alpha_3 = 0.93a_0$, $\alpha_5 = 0.90a_0$, where $a_0 = \sqrt{\phi_0/H_{c_2}}$. The effective length $R_e = \ell\sqrt{(1+\tau/2)/2}$, where $\tau = 2\cos(\pi/5)$ is the golden mean. The expression for $\Delta G$ is valid only when $H_{c_2} \gg \phi_0/R_e^2$. The second term in $\Delta G$ takes into account the boundary mismatch energy, and indicates that as the size of the decagon cylinder gets smaller it becomes more favorable to have a 5-fold quasi-crystal rather than a regular triangular lattice. Notice, however, that when $R_e \to \infty$ the triangular lattice has lower Gibbs free energy as it must, and no transition to a 5-fold quasi-crystal occurs. Thus, this possible transition may occur for finite samples only. From the condition that $\Delta G = 0$ it is easy to obtain

$$H_Q = H_{c_2}\left[1 - \beta^*\kappa^*\sqrt{\frac{(\alpha_3-\alpha_5)}{R_e}}\right] \qquad (10)$$

where the transition to a quasi-crystal occurs. Here, $\kappa^* = \sqrt{2\kappa^2-1}/\kappa$. The phase diagram for a superconductor with $\kappa = 20$, $H_{c_2}(0) = 10T$ and $R_e = 10^{-6}m$ is shown [7] below using $H_{c_2}(T) = H_{c_2}(0)\left[1 - (T/T_c)^2\right]$.

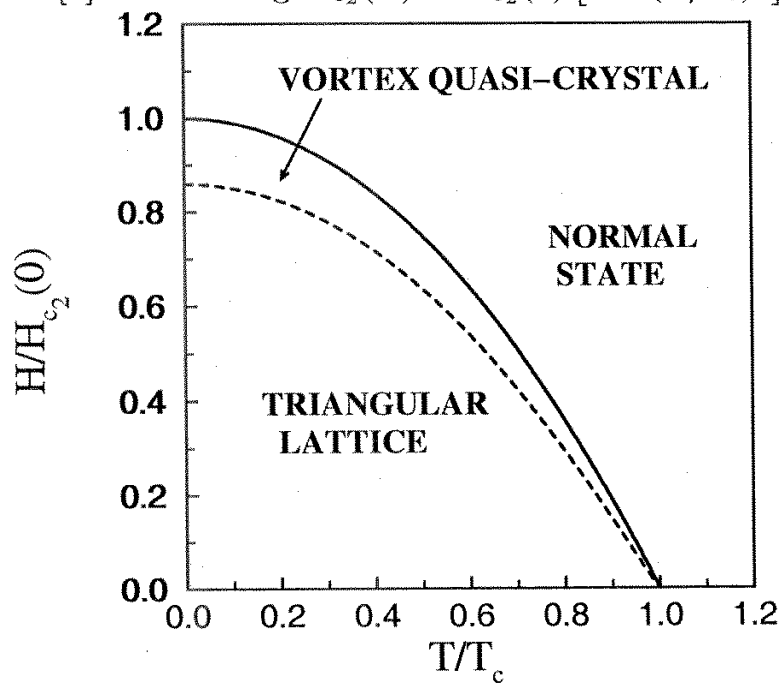

FIG. 1. H vs. T phase diagram for a long decagon cylinder cut out of a regular cylinder of radius for $R_e = 10^{-6}m$. The superconductor is assumed to have $\kappa = 20$, $H_{c_2}(0) = 10T$.

The jump discontinuity of the magnetization $\Delta M (= M_3 - M_5)$ as a function of temperature at the critical field $H_Q$ can be calculated from the Gibbs free energy leading to

$$\Delta M = \frac{H_Q}{4\pi}\left[\Delta M_b + \Delta M_s\right], \qquad (11)$$

where $\Delta M_b = -\epsilon/(2\kappa^2-1)$ and $\Delta M_s = (\epsilon/\kappa^*)^2/(2\kappa^2)$, with $\epsilon = (\gamma^*/\beta^*)/(1-\gamma^*)$ and $\gamma^* = \beta^*\kappa^*\sqrt{(\alpha_3-\alpha_5)/R_e}$. A plot of $\Delta M$ is illustrated in Fig. 2 for the same parameters of Fig. 1. Using these parameters produces jump discontinuities $\Delta M \approx -0.62G$ at $T = 0$, and $\Delta M \approx -0.29G$ at $T = 0.8T_Q$. Notice that $\Delta M < 0$ indicates that the 5-fold vortex quasi-crystal is denser than the triangular vortex lattice at $H_Q$, being at best a few percent denser at $T = 0$.

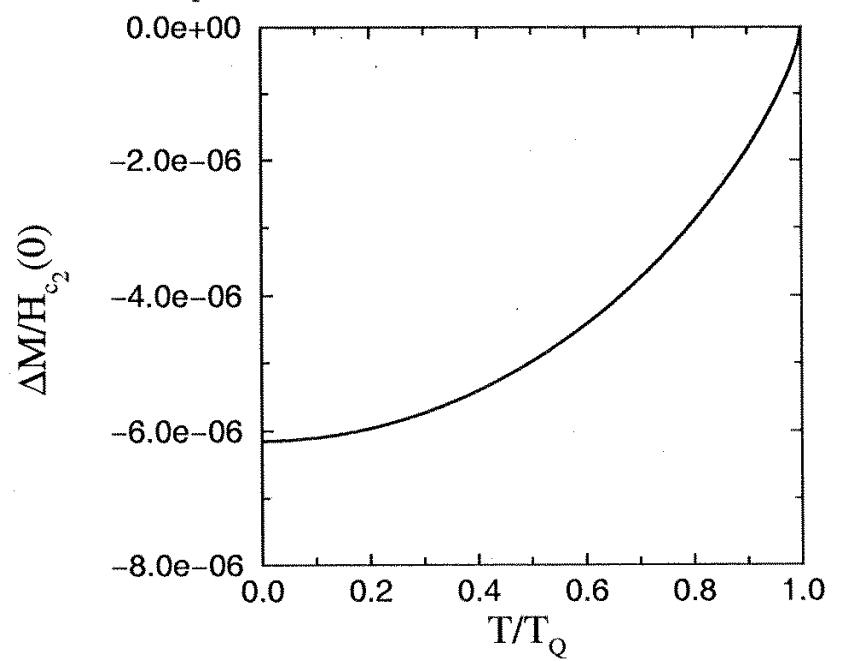

FIG. 2. The jump discontinuity $\Delta M = M_3 - M_5$ at the critical field $H_Q$ for various reduced temperatures $T/T_Q$, for the same parameters of Fig. 1.

In addition to magnetization measurements, calorimetric experiments are also interesting to look at. However, specific heat measurements are hard to perform because of the large sample requirement, and since the sample size is important for the present discussion it is not clear that such experiment can be succesfully performed. Nevertheless, the thermodynamic relationship between the magnetization and entropy jumps is revealed in the Clapeyron equation $\Delta S = -\Delta M dH_Q/dT$. Since $dH_Q/dT < 0$, and $\Delta M < 0$ implies that $\Delta S < 0$, i.e, the entropy $S_3$ of the triangular vortex lattice is less than the entropy $S_5$ of the 5-fold vortex quasi-crystal indicating that latent heat $L = T\Delta S$ is required to cause this phase transition.

Thermodynamic quantities can provide a good understanding of properties averaged over the entire sample. The use of local probes, however, is much desired in order to reveal the change in structure from a triangular vortex crystal to a 5-fold vortex quasi-crystal. Thus, neutron scattering, Bitter decoration and scanning tunneling microscopy (STM) experiments can help ellucidate the structure of the vortex arrangement.

In neutron diffraction experiments periodic or quasi-periodic variations of $\mathcal{H}(x,y)$ will result in Bragg peaks. The position of these peaks thus determine the characteristic length scale of the vortex structure and its symmetry. The neutron scattering amplitude in the Born approximation is

$$b(\mathbf{q}) = \frac{M_n}{2\pi\hbar^2} \int \mu_n H(\mathbf{r}) \exp(i\mathbf{q} \cdot \mathbf{r}) d\mathbf{r}, \quad (12)$$

where $\mu_n = 1.91 e\hbar/M_n c$ is the neutron magnetic moment and the $M_n$ is the neutron mass. The scattering amplitude $b(\mathbf{q})$ is directly proportional to the Fourier transform $H(\mathbf{q})$ ($\mathcal{H}(q_x, q_y)$) of the microscopic field $H(\mathbf{r})$ ($\mathcal{H}(x,y)$) of eqn. 2. The neutron scattering cross section $\sigma(q_x, q_y) = 4\pi^2 |b(q_x, q_y)|^2$ has sharp peaks at $(q_x, q_y) = (0,0)$ (central peak) and at $(q_x, q_y) = (\pm q_{xNm}, \pm q_{yNm})$ (first Bragg peaks), where $q_{xNm} = Q_N \cos(m\pi/N)$ and $q_{yNm}) = Q_N \sin(m\pi/N)$, with $m = 0, 1, ..., N-1$. For the triangular lattice $N = 3$ the first Bragg peak occurs at $|Q_3| = 2.31 \times \pi/d_3$, where $d_3$ is the lattice spacing. For the 5-fold vortex quasi-crystal (Penrose Lattice) $N = 5$ the first Bragg peak occurs at $|Q_5| = 2.46 \times \pi/d_5$, where $d_5$ is the side of a tile. Since the sample size is important for the observation of a 5-fold quasi-crystal, neutron scattering experiments may be difficult to perform. Thus, Bitter decoration or STM may be better techniques. For instance, STM scans at different fields and temperatures in the vicinity of $H_Q(T)$ should reveal the real space locations of vortices, which can be Fourier transformed (FT) to obtain a a 6-fold pattern for the triangular vortex lattice and a 10-fold pattern for the 5-fold vortex quasi-crystal (Penrose lattice), see Fig. 3.

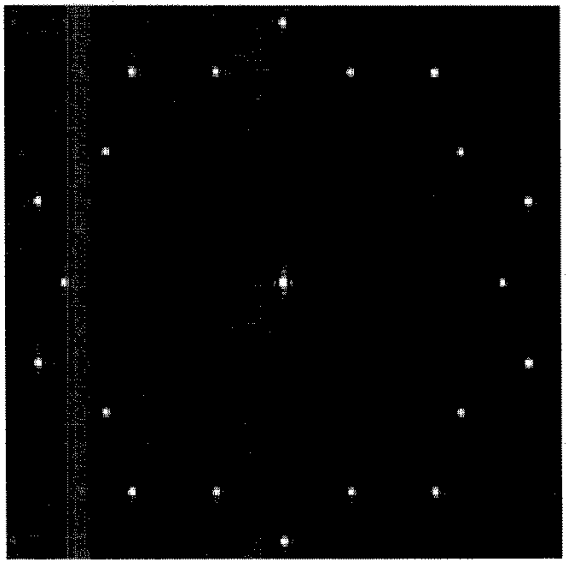

FIG. 3. First peaks of the square of the FT pattern for the 5-fold vortex quasi-crystal (Penrose lattice). Notice the 10-fold symmetry revealed.

Now that the phase diagram, thermodynamics, and the local signatures of a vortex-crystal have been discussed, it is important to say a word on the stability of such structures. An stability analysis in the free energy can be performed by moving the vortices away from their equilibrium positions $z_i$ to $z_i + \delta z_i$. The eigenvalues associated with these displacements indicates that for $H > H_Q(T)$ the vortex quasi-crystal lattice is stable. However, there is no full compatibility of the 5-fold vortex lattice with the decagon cylinder geometry, and thus the appearance of disclinations and dislocations is possible.

In conclusion, this initial theoretical investigation has shown that vortex quasi-crystals may be experimentally observed in isotropic type II superconductors, provided that the sample size and shape are properly chosen. However, it is an open theoretical and experimental question whether vortex quasi-crystalline phases exist in the context of layered superconductors and, in particular, high-temperature superconductors, where vortex liquid [8] and vortex glass [9] phases have been experimentally observed. An extrapolation of the present work suggests that by taking into account boundary effects, sample shape and size in high-temperature superconductors (layered superconductors), a vortex quasi-crystal phase may exist and compete with vortex liquid [10,11], vortex glass [12] or vortex crystal [5] phases, as magnetic field, temperature and disorder are varied.

Many thanks to the Aspen Center for Physics, and to the NSF (DMR 9803111) for partial support.

Dr. C.A.R. Sá de Melo
School of Physics
Georgia Institute of Technology
Atlanta, GA 30332

The Editors are considering publication of the manuscript by:

C.A.R. Sá de Melo

entitled:

Vortex quasicrystals

It is understood that this manuscript has not been published and is not under active consideration for publication elsewhere. Submission of this manuscript implies acceptance by the authors of the established procedures for selecting manuscripts for publication.

Information concerning the processing of this manuscript will be sent as soon as possible.

We no longer routinely return manuscripts with correspondence. You may request return of the manuscript and/or figures.

With each resubmission, please state if figures have changed or not, and if the scanner-reproducible-quality versions are included, being sent, or already in our hands.

EDITORIAL NOTES:
( ) Please see attached
( ) See revised title
( )

Please notify us immediately if there is an error in authors, title, or correspondence address.

At your convenience, we would appreciate receipt of items checked below:

( ✓ ) Properly executed copyright transfer form made out to the American Physical Society.

( ) Communications information (if available)

Telephone ______________________

Fax ______________________

Internet ______________________

length_est_prl.tex
revised 2/99

PHYSICAL REVIEW LETTERS

Manuscript No. LH7266

Date 8-16-99

First Author de Melo

We estimate the length of the manuscript as shown to the right. Guidelines for estimating the length are given at the end. [Manuscripts that we estimate to be within our page limits according to a quick count formula are marked "O.K." and are not given a line count.]

If the designation "O.K." is given, or the line count is 460 or less, there should be no problem with the manuscript's length. Note that any subsequent increase in the length may cause the manuscript to exceed the page limit and, if the manuscript is accepted, result in a request for cuts to the paper at a later stage.

**Warning:** If the count is between 460 and 500 (or would be with relettered figures) there is a greater uncertainty that the length limit of four journal pages will be met.

- ☐ Your manuscript is being sent to the referees but a shortened version ($\leq 460$) will be required if the paper is accepted.
- ☐ Your manuscript is too long; shortening to $\leq 460$ is required before further processing.

**Note:** The 460-line restriction also applies to any subsequent version of the manuscript; care should be taken when making modifications to stay within the allowed length. Revised manuscripts sent by conventional mail should be submitted in quadruplicate.

LINE COUNT (PRL one-column lines)

Title — 4

Bylines — 4

Receipt date — 6

Abstract — 16

PACS numbers — 4

Text (____characters/avg.author line) × (332 author lines)/55= 332

Equations

References — 30

Tables and Captions

Figure Captions — 18

Figures (6 lines/inch) — 44

1. 1col. 14
2. 1col. 14
3. 1col. 16

TOTAL 458

( )RR  OK

For electronically submitted REVTEX compuscripts, the length estimate is based on the galley-format output. Note that page-composition requirements may add additional spacing.

Please supply any items checked below. Supplemental information may be enclosed.

( ) Original printed manuscript or clear black-on-white reproduction (printed on one side of page).

( ) Double-spaced (no more than 3 lines/in.) abstract, references, captions, manuscript.

( ) Abstract of no more than 600 characters including spaces.

( ) Original india-ink drawings or prints suitable for scanner-reproduction.

( ) Copy of table suitable for scanner-reproduction.

( ) Table and figure captions.

( ) Byline changes  Required ☐  Optional ☐

( ) Pages with equations printed or handwritten in ink, not pencil, of size and legibility that the symbols can be distinguished by a production keyboarder.

( ) A manuscript copy in which the type size is larger ($\lesssim 90$ characters/spaces per 6-in. line), for use in the composition process.

FIGURE CHANGES  Required ☐  Optional ☐

( ) Numbered figures (Fig. 1, Fig. 2, etc.)

( ) Original Figure(s) with

_____ larger lettering and/or data symbols (relative to size of figure itself) so that figure may be scanner-reproduced at a ___-column width (for a savings of _____ PRL lines). In order to reproduce figure(s) _____ as indicated above, subscripts and superscripts must be enlarged to _____ in height and all other lettering _____ in height.

_____ finer and/or larger lettering to prevent its filling in when figure is scanner-reproduced. Minimum lettering size for the enclosed figure is _____.

_____ lettering and lines that are not blurred, faded, broken, or smudged.

_____ uniform lettering size.

( ) Use of bylines and footnotes conforming to PRL style (see recent issue and/or enclosed memo) will save _____ lines.

ADDITIONAL COMMENTS:

Enclosures:  ☐ D Memo  ☐ Manuscript Type Size  ☐ Author's Original Figure  ☐ Byline Memo

JD